\PassOptionsToPackage{unicode}{hyperref}
\PassOptionsToPackage{hyphens}{url}
\documentclass[
]{article}
\usepackage{amsmath,amssymb}
\usepackage{lmodern}
\usepackage{iftex}
\ifPDFTeX
  \usepackage[T1]{fontenc}
  \usepackage[utf8]{inputenc}
  \usepackage{textcomp} 
\else 
  \usepackage{unicode-math}
  \defaultfontfeatures{Scale=MatchLowercase}
  \defaultfontfeatures[\rmfamily]{Ligatures=TeX,Scale=1}
\fi
\usepackage{multirow}

\makeatother
\usepackage{xcolor}
\usepackage[margin=1in]{geometry}
\usepackage{graphicx}
\makeatletter
\def\maxwidth{\ifdim\Gin@nat@width>\linewidth\linewidth\else\Gin@nat@width\fi}
\def\maxheight{\ifdim\Gin@nat@height>\textheight\textheight\else\Gin@nat@height\fi}
\makeatother
\setkeys{Gin}{width=\maxwidth,height=\maxheight,keepaspectratio}
\makeatletter
\def\fps@figure{htbp}
\makeatother
\usepackage{xspace}

\usepackage[backend=bibtex,
maxbibnames=6,
minbibnames=6,
mincitenames = 1,
maxcitenames = 1,
citestyle = authoryear,
bibstyle = authoryear,
sortcites=false,
useeditor=false,
url=false,
eprint=false,
sorting = nyt,
isbn=false,
doi=false,
giveninits=true,
uniquename = init,
citetracker = true
]{biblatex}
\AtEveryCitekey{\ifciteseen{}{\defcounter{maxnames}{3}}}
\AtEveryBibitem{
\clearname{editor}
\clearfield{pmid}
\clearfield{note}
\clearfield{issn}
}

\usepackage{grffile}
\usepackage{amsbsy}
\usepackage{bbold}
\usepackage{color}
\usepackage{amsmath}
\usepackage{dsfont}
\usepackage{nicefrac}
\usepackage{tikz}
\usetikzlibrary{matrix,shapes,arrows,positioning,chains}
\usepackage{xcolor}
\usepackage{caption}
\usepackage[utf8]{inputenc}

\newcommand{\bs}[1]
   {\boldsymbol{#1}}
\newcommand{\mb}[1]
   {\mathbf{#1}}

\newcommand{\Cov}[1]
  {\text{Cov}(#1)}
  \newcommand{\Cor}[1]
  {\text{Cor}(#1)}
\newcommand{\Var}[1]
  {\text{Var}(#1)}

\newcommand*\fref[1]{Figure \ref{#1}}
\newcommand*\sfref[1]{Supplementary Figure \ref{#1}}

\newcommand{\rsq}
  {\(R^2\)\xspace}

\newcommand{\hrsq}
    {\(\hat{R^2}\)\xspace}

\usepackage{xr-hyper}
\usepackage{hyperref}
\usepackage{url}
\usepackage{caption}
\makeatletter
\long\def\XR@test#1#2#3#4\XR@{%
  \ifx#1\newlabel
    \mod@newlabel{\XR@prefix#2}#3{}{}\@nil
  \else\ifx#1\@input
     \edef\XR@list{\XR@list#2\relax}%
  \fi\fi
  \ifeof\@inputcheck\expandafter\XR@aux
  \else\expandafter\XR@read\fi}
\def\mod@newlabel#1#2#3#4\@nil{\newlabel{#1}{{#2}{#3}}}
\makeatother
\usepackage{gensymb}
\usepackage{longtable}
\IfFileExists{xurl.sty}{\usepackage{xurl}}{} 
\hypersetup{
  pdftitle={The out-of-sample \rsq},
  pdfauthor={Stijn Hawinkel, Willem Waegeman and Steven Maere},
  hidelinks,
 pdfcreator={LaTeX via pandoc}}
  
\newcommand{\blind}{0}

\begin{document}

\def\spacingset#1{\renewcommand{\baselinestretch}%
{#1}\small\normalsize} \spacingset{1}

\if0\blind
{
  \title{\bf The out-of-sample \rsq: estimation and inference}
  \author{Stijn Hawinkel\thanks{
    S.H gratefully acknowledges funding by Inari Agriculture NV, W.W. received funding from the Flemish
Government under the “Onderzoeksprogramma Artifici\"{e}le
Intelligentie (AI) Vlaanderen”.}\hspace{.2cm}\\
    Department of Plant Biotechnology and Bioinformatics, Ghent University\\
    VIB Center for Plant Systems Biology\\
    and \\
    Willem Waegeman \\
    Department of Data Analysis and Mathematical Modelling, Ghent University\\
    and\\
    Steven Maere\\
    Department of Plant Biotechnology and Bioinformatics, Ghent University\\
    VIB Center for Plant Systems Biology\\    
    }
  \maketitle
} \fi

\if1\blind
{
  \bigskip
  \bigskip
  \bigskip
  \begin{center}
    {\LARGE\bf Title}
\end{center}
  \medskip
} \fi

\bigskip
\begin{abstract}
Out-of-sample prediction is the acid test of predictive models, yet an
independent test dataset is often not available for assessment of
the prediction error. For this reason, out-of-sample performance is
commonly estimated using data splitting algorithms such as
cross-validation or the bootstrap. For quantitative outcomes, the ratio
of variance explained to total variance can be summarized
by the coefficient of determination or in-sample \rsq, which is easy to interpret
and to compare across different outcome variables. As opposed to the in-sample \rsq, the out-of-sample \rsq has not been well defined and the variability on the out-of-sample \hrsq has been largely ignored. Usually only its point
estimate is reported, hampering formal comparison of predictability of
different outcome variables. Here we explicitly define the out-of-sample \rsq as a comparison of two predictive models, provide an unbiased estimator and exploit recent theoretical advances
on uncertainty of data splitting estimates to provide a standard error
for the \hrsq. The performance of the estimators for the \rsq and its standard error are investigated in a simulation study. We demonstrate our new method by constructing confidence
intervals and comparing models for prediction of
quantitative \emph{Brassica napus} and \emph{Zea mays} phenotypes based
on gene expression data.
\end{abstract}

\noindent%
\vfill

\newpage
\spacingset{1.45} 
\section{Introduction}
\label{sec:intro}

Predictive model performance is evaluated through loss functions, which
quantify the discrepancy between predicted and observed values. For
quantitative outcomes, the most popular loss function is the squared
error loss, defined as
\(l(y_i, m(\mb x_i)) = (y_i-m(\mb x_i))^2\), with
\(y_i\) the observed outcome in sample \(i\)=1,\ldots,\(n\), with \(n\)
the sample size, and \(m(\mb x_i)\) the outcome predicted by a model $m$ using a set of predictors \(\mb x_i\). Most often, summary statistics of
the loss distribution are reported, such as the mean squared error
(MSE), which is estimated as
\(n^{-1}\sum_{i=1}^nl(y_i, m(\mb x_i))\) \parencite{Hastie2009}.
Yet unlike misclassification loss for categorical outcomes, the MSE is
not easy to interpret, as it depends on the degree of variability of the
dataset under study as well as on the measurement unit used. For this
reason, the MSE is often compared to the total variance in the sample:
\(n^{-1}\sum_{i=1}^nl(y_i, \bar{y})\) with $\bar{y}$ the average outcome, here called mean of squares
total (MST) in analogy to the MSE. This gives rise to the coefficient of
determination or \rsq, which is defined as \parencite{Kvalseth1985}

\begin{equation}
R^2 = 1-\frac{MSE}{MST}
\label{eq:R2in}
\end{equation}

The \rsq is unitless and hence comparable across outcome variables with
different units \parencite{Valbuena2019}. It is often employed as a
goodness-of-fit statistic of a model to a given sample. In this case,
values $m(\mb x_i)$ and $\bar{y}$ are predicted using a model trained on the entire
sample, so including observation \(i\), and the \rsq ranges from 0 to 1
and can be interpreted as the proportion of variance explained by the
model \parencite{AndersonSprecher1994}. Modifications have been proposed
to penalize for model complexity of general linear models, e.g.~the
adjusted \rsq \parencite{Wherry1931}, to better guide model building.
The \rsq has also been extended to generalized linear models
\parencite{Zhang2017, Cameron1997, Nagelkerke1991} and survival analysis
\parencite{Verweij1993}. For linear models, a standard error and
confidence intervals have been derived for in-sample
\rsq \parencite{Cohen2014}.

Yet these \rsq values are intended as goodness-of-fit
diagnostics and model building aides \emph{within} a single dataset. Modern prediction models, e.g. random forests \parencite{Breiman2001}, are more
flexible and make less assumptions about the type of association between
outcome and predictors than linear models. Moreover, linear models may be applied to high-dimensional omics datasets with many more predictors than observations. Since both these scenarios may
cause overfitting, in this case the in-sample loss is a poor measure for
predictive performance, which is instead evaluated on data not used to
train the model. Ideally this is an independent test dataset, but more
often the out-of-sample loss is estimated on the same dataset using data
splitting algorithms such as cross-validation (CV) \parencite{Bates2021}
or the bootstrap \parencite{Efron1997}. Thereby the loss is estimated on
a different part of the dataset than was used for building the model. As
a consequence, the aforementioned results on the in-sample \rsq and its standard error and
confidence interval no longer hold. The out-of-sample \rsq values for instance lie in the interval \(]-\infty, 1]\), instead of [0, 1] for the in-sample \rsq. Still, for
statistical inference, a formal definition of the out-of-sample \rsq statistic is needed, as well as some measure of uncertainty on the point estimate \(\hat{R}^2\). Here we define the out-of-sample \rsq as a model
comparison of the prediction model at hand with a baseline prediction
model that ignores covariate information, provide an unbiased estimator for it and exploit recent advances in the field of data splitting algorithms \parencite{Bates2021} to present a standard error (SE) for
the out-of-sample \hrsq. We validate the estimators for the out-of-sample \rsq and its standard error in a simulation study. Subsequently, we illustrate how these estimators can be used for comparing predictability of outcome variables and for building confidence intervals on real omics datasets of \emph{Brassica napus} and
\emph{Zea mays} field trials. 

\section{The out-of-sample \rsq: definition and inference}\label{results}

\subsection{The out-of-sample R² as a model comparison}\label{sec:oosR2}

We propose to regard the out-of-sample \rsq as a comparison of two
out-of-sample prediction models, just like the regular in-sample \rsq is
a comparison of two in-sample models
\parencite{AndersonSprecher1994, Campbell2008}. Call $m_d$ a prediction model that is trained on a dataset \textbf{d} = ($\mathbf{y}$, $\mathbf{x}$), which can be used to make predictions for a new y given $\mathbf{x}$: $m_d(\mb{x})$. To score these prediction models $m$, we are interested in the expected squared error loss for a hypothetical out-of-sample observation $Y_{oos}$ with respect to its predicted value $m_D(\mb{X}_{oos})$, averaged over all possible datasets \textbf{D} = (\textbf{Y}, \textbf{X}) of fixed size $n$ on which $m$ was estimated. More formally, we are looking for

\begin{equation}
E_{\mathbf{D}}\left[E_{(Y, \mb X)_{oos}}\left(\left[Y_{oos}-m_D(\mb{X}_{oos})\right]^2\mid \mathbf{D}\right)\right],
\label{eq:Err}
\end{equation}

with the inner expectation running over all out-of-sample observations and the outer expectation over all datasets. We work under the common scenario where no independent test set of out-of-sample observations is available, so all estimation needs to happen based on a single observed dataset \textbf{d} of size $n$. Here and in what follows, all variables without subscript $oos$ belong to \textbf{d}.

The first model in the comparison, referred to as the null model \parencite{AndersonSprecher1994}, simply
uses the average outcome of the observed data $\bar{Y}$ as prediction for all
out-of-sample observations, ignoring available predictors. The expected loss of this model is the out-of-sample MST can be estimated analytically from the vector of outcome values \textbf{y} as derived below, relying on the equality E($Y$) = E($\bar{Y}$) = E($Y_{oos}$):

\begin{equation}
\begin{aligned}
MST &= E_{\mathbf{D}}\left[E_{Y_{oos}}\left(\left[Y_{oos}-\bar{Y}\right]^2\mid \mathbf{D}\right)\right]\\
&=E_{\mathbf{D}}\left[E_{Y_{oos}}\left(\left[(Y_{oos}-E(Y)) + (E(Y)-\bar{Y})\right]^2\mid \mathbf{D}\right)\right]\\
&=E_{\mathbf{D}}\left[E_{Y_{oos}}\left(\left[Y_{oos}-E(Y)\right]^2  + \left[E(Y)-\bar{Y}\right]^2 + 2(Y_{oos}-E(Y))(E(Y)-\bar{Y})\mid \mathbf{D}\right)\right]\\
&=E_{\mathbf{D}}\left[\Var{Y}  + \left(E(\bar{Y})-\bar{Y}\right)^2 + 2 E_{Y_{oos}}\left((Y_{oos}-E(Y))(E(Y)-\bar{Y})\mid \mathbf{D}\right)\right]\\
&=E_{\mathbf{D}}\left[\Var{Y}  + \left(E(\bar{Y})-\bar{Y}\right)^2 + 2 (E(Y)-\bar{Y})E_{Y_{oos}}\left(Y_{oos}-E(Y_{oos})\mid \mathbf{D}\right)\right]\\
&= \Var{Y} + \Var{\bar{Y}} +0 = \frac{n+1}{n}\Var{Y}.
\end{aligned}
\label{eq:MST}
\end{equation}

This result nicely illustrates how the expected loss is a sum of variability
of the estimator around the expected value and the variability of the
observations around the expected value. An unbiased estimator for the population
variance \(\Var{Y}\) is provided by
\((n-1)^{-1}\sum_{i=1}^n(y_i-\bar{y})^2\). The estimator for the out-of-sample
MST then inflates this estimator by a factor (n+1)/n to account for the
variability in the estimation of E(Y) through $\bar{Y}$.

The second model in the comparison is the prediction model that makes use of the covariate information. Since it is a more complicated model, no analytical expression for its expected out-of-sample loss (the MSE) is available,
so that, for want of independent test data, it needs to be estimated through data splitting algorithms. Here we discuss cross-validation and the .632 bootstrap \parencite{Efron1997}, but other options are possible, e.g. a single split of the available data in a training and a test dataset.

In cross-validation (CV), the samples of $\mb d = (\mb y, \mb x)$ are divided into $K$ equally sized folds of $n/K$ observations, assuming for simplicity that $K$ divides $n$. The set of samples in fold $k = 1, \hdots, K$ is indicated by $\mb d_k$, the other observations as $\mb d_{\setminus k}$. For all $k$, the model is trained on $\mb d_{\setminus k}$ yielding model $m_k$. The squared error loss of this fold is estimated as $\frac{K}{n}\sum_{i \in \mb d_{k}}(y_i-m_k(\mb x_i))^2$, and the overall estimate of the MSE becomes $n^{-1}\sum_{k=1}^K\sum_{i \in \mb d_k}(y_i-m_k(\mb x_i))^2$. Repeating the splitting into folds reduces the variability of the estimate $\widehat{MSE}$; the final estimate is then the average $\widehat{MSE}$ over the repeated splits. The procedure outlined above is what we refer to as simple CV. For nested CV, we refer to \textcite{Bates2021}, who provide an estimate of $\Var{\widehat{MSE}}$, as well as a correction for the fact that $m_k$ is estimated on a dataset of size $n\frac{K-1}{K}$ rather than $n$.

The .632 bootstrap is an alternative way of estimating the MSE \parencite{Efron1997}. In this case, the samples are split by resampling $n$ samples with replacement. A sample has a probability of $1-\exp(-1) \approx 0.632$ of being contained in this bootstrap sample, hence the name. This resampling step is repeated B times, with $\mb d_b$ indicating the set of included samples (possibly containing the same sample several times), $m_b$ the model trained on this set and $\mb d_{\setminus b}$ the set containing the unique excluded samples, with $b=1,\hdots,B$. Call $N_i^b$ the number of times sample $i$ is included in $\mb d_b$, and $J_i^b = I(N_i^b=0)$ with I() the indicator function. The .632 bootstrap estimate of the MSE is then given by $n^{-1}\exp(-1)\sum_{i=1}^n\left((y_i-m_d(\mb x_i))^2\right) + n^{-1}(1-\exp(-1))\sum_{i=1}^n\left[\frac{\sum_{b=1}^B\left(J_i^b(y_i-m_b(\mb x_i))^2)\right)}{\sum_{b=1}^BJ_i^b}\right]$, with $m_d$ trained on the set of all samples. The bootstrap .632 estimate is thus a weighted average of in-sample and out-of-sample error, but can be written as a weighted sum over all samples. A standard error on the .632 estimate, as well as variations on this estimator, are provided by \textcite{Efron1997}. 
 \textcite{Bates2021} show that CV and the .632 bootstrap indeed estimate the quantity in \eqref{eq:Err}; CV is known to be unbiased whereas the bootstrap is slightly biased for MSE estimation \parencite{BragaNeto2004, Jiang2007, Kohavi2001, Molinaro2005}. 

The out-of-sample \rsq is then a population parameter defined as

\begin{equation}
R^2 = 1-\frac{E_{\mathbf{D}}\left[E_{Y_{oos}}\left(\left[Y_{oos}-m_D(\mathbf{X}_{oos})\right]^2\mid \mathbf{D}\right)\right]}{E_{\mathbf{D}}\left[E_{Y_{oos}}\left(\left[Y_{oos}-\bar{Y}\right]^2\mid \mathbf{D}\right)\right]},
\label{eq:R2oos}
\end{equation}

and depends on the sample size $n$ of the data $\mathbf{D}$ on which the prediction model is trained, on the prediction model and on the joint distribution of the outcome and predictors. The value of the out-of-sample \rsq reflects a comparison of the null model with the more elaborate prediction model: when it is smaller than 0, the null model achieves the best out-of-sample predictions; when it is larger than 0, the elaborate prediction model achieves the lowest out-of-sample squared error loss. In the latter case, the \rsq can be interpreted as the proportion of the null model's squared error loss explained by the elaborate model. For estimation of the out-of-sample \rsq, we plug in the aforementioned estimators of the out-of-sample squared errors MST and MSE based on the observed data:

\begin{equation}
\hat{R}^2 = 1-\frac{\widehat{MSE}}{\widehat{MST}} = 1-\frac{\widehat{MSE}}{(n+1)/(n(n-1))\sum_{i=1}^n(y_i-\bar{y})^2}.
\label{eq:R2cv}
\end{equation}

This expression can be seen as the prediction equivalent of the forecasting out-of-sample
\rsq by \textcite{Campbell2008}.

\subsection{The pooling and averaging estimators for the \rsq}\label{sec:estR2}

In equation \eqref{eq:R2cv} above, the squared error losses are calculated and summed over all observations before combining them into the final \rsq estimate. This is called the \emph{pooling} strategy in the field of multivariate loss function estimation (e.g. area-under-the-curve (AUC) estimation), as opposed to the \emph{averaging} approach whereby the performance measure is calculated for every left-out fold separately, and then averaged over the folds to obtain the final estimate \parencite{Bradley1997}. The latter approach is not applicable for .632 bootstrap estimation of the MSE, but can be employed when using cross-validation. In this case, the MSE and MST are estimated for every left-out fold separately as follows. The MSE is estimated per fold as in Section \ref{sec:oosR2}. For the MST, we consider the following two estimation strategies: either as in \eqref{eq:MST} based on the training folds, adapting $n$ to be the sample size of the training folds, or as the empirical variance of the left-out fold $\frac{1}{\left[\sum_{i=1}^nI(i \in \mb d_k)\right]-1}\sum_{i \in \mb d_k} (y_i-\bar{y}_{\mb d_k})^2$ with $\bar{y}_{\mb d_k}$ the average of the left-out fold $\mb d_k$ and $I()$ the indicator function \parencite{Valbuena2019}. In both cases, these estimates are combined into an \rsq estimate for every fold separately and then averaged. We call the two presented estimators "averaging \rsq with training MST" and "averaging \rsq with test MST", respectively.

\subsection{\texorpdfstring{The standard error of \hrsq
\label{sec:sersq}}{The standard error of \hrsq}}\label{a-standard-error-for-ruxb2}

As a measure of uncertainty on the estimate \(\hat{R}^2\), we derive an
expression for its standard error (SE) $\left(\text{SE}(\hat{R}^2)  = \Var{\hat{R}^2}^{1/2}\right)$. For this, we rely on asymptotic
normality of the estimator \hrsq. According to the first order delta method
\parencite{Gauss1823},

\begin{equation}
\Var{\hat{R}^2} \approx \left(\nabla R^2\right)^T \begin{bmatrix} \Var{\widehat{MSE}} & \Cov{\widehat{MSE}, \widehat{MST}}\\ \Cov{\widehat{MSE}, \widehat{MST}} & \Var{\widehat{MST}} \end{bmatrix} \nabla R^2.
\label{eq:delta}
\end{equation}

The gradient is found as
\(\left(\nabla R^2\right)^T = \left(\frac{\partial R^2}{\partial MSE}, \frac{\partial R^2}{\partial MST}\right) = \left(-\frac{1}{MST},\frac{MSE}{MST^2}\right)\),
and is evaluated at the estimates \(\widehat{MSE}\) and
\(\widehat{MST}\). An estimate of \(\Var{\widehat{MSE}}\) for
cross-validation estimation of the MSE was provided by
\textcite{Bates2021}, and for .632 bootstrap estimation of the MSE by
\textcite{Efron1997}. The estimate of \(\Var{\widehat{MST}}\) is given
by \(\frac{2}{(n-1)}\widehat{MST}^2\) \parencite{Harding2014}. 

The covariance \(\Cov{\widehat{MSE}, \widehat{MST}}\) is usually
considerable and positive, since the MSE and MST are estimated on the
same outcome vector \textbf{y}. It can be decomposed as
\(\Cov{\widehat{MSE}, \widehat{MST}} = \Cor{\widehat{MSE}, \widehat{MST}}\left(\Var{\widehat{MSE}}\Var{\widehat{MST}}\right)^{1/2}\).
\(\Cor{\widehat{MSE}, \widehat{MST}}=\rho\) cannot be derived analytically, such that it is estimated either
via nonparametric or parametric bootstrap, or via jackknifing as follows. Bootstrap samples of size $n$ are either drawn nonparametrically, by sampling entries with replacement from the data \textbf{d}, or parametrically by assuming a parametric model, estimating the corresponding parameters from the data and drawing \textbf{Y} from the model parameterized by these estimates while keeping \textbf{x} fixed. The MSE and MST are then estimated for every bootstrap sample in the same way as for the original sample. In jackknifing, each observation is dropped in turn, and the MSE and MST are estimated on the remaining samples of size $n-1$, leading to a number of jackknife estimates equal to the sample size $n$. For the bootstrap and jackknife samples, simple rather than nested CV is used for MSE estimation to reduce computation time, but the splitting into CV folds is repeated as for the MSE estimation on the observed dataset. The empirical
correlation between the bootstrap or jackknife estimates of the MSE and MST is used as estimate of the correlation between the estimators for MSE and MST. When using the .632 bootstrap for MSE estimation in combination with
bootstrapping for the estimation of the correlation between MSE and MST,
we refer to the former as inner bootstrap samples and to the latter as
outer bootstrap samples.

\subsection{Inference on the \rsq}

Given the standard error estimate, one-sided approximate z-tests can be used to test the null hypothesis that \rsq $\leq 0$, i.e. whether the prediction model is significantly better than the null model. Another popular application of standard errors is the construction of
confidence intervals, again relying on asymptotic normality of the
estimator \hrsq. The confidence interval is then constructed as

\begin{equation}
\hat{R}^2 \pm \widehat{\text{Var}}({\hat{R}^2})^{1/2}\Phi^{-1}\left(\frac{\alpha}{2}\right)
\label{eq:confInt}
\end{equation}

with \(\alpha\) the significance level and \(\Phi^{-1}\) the inverse cumulative
distribution function of the standard normal distribution. The upper
bound of the confidence interval is truncated at 1.

As an alternative to the delta method SE from \eqref{eq:delta}, the
standard deviation of \rsq estimates across nonparametric or parametric bootstrap samples (obtained in the same way as for estimating $\rho$) can be used as an estimate of the SE (further referred to as the bootstrap SE). As alternatives for the confidence intervals based on the delta method SE, we consider percentile and bias-corrected
and accelerated (BCa) bootstrap confidence intervals constructed
based on the distribution of the bootstrap \rsq estimates
\parencite{DiCiccio1996}, as well as confidence intervals constructed from the bootstrap SE in the same way as for the delta method SEs in \eqref{eq:confInt}. This
latter method was chosen rather than the bootstrap-t method
\parencite{DiCiccio1996} since this would require calculating standard
errors for every bootstrap instance and hence be too computationally
intensive.

\section{Simulation study}\label{simulation-study}

We conduct a simulation study in which we apply the proposed methodology on \rsq estimation on a one-dimensional prediction model (OLS) and a high-dimensional prediction model (EN). We study the performance of the estimators for the MST and MSE, and compare the pooling and averaging estimators for the \rsq. Next we compare our delta method estimator for the SE on the \hrsq and corresponding confidence intervals with competitor methods based on the bootstrap SE and percentile and BCa confidence intervals.

\subsection{\texorpdfstring{Setup
\label{sec:setup}}{Data generation }}\label{data-generation}

\subsubsection{Data generation}\label{sec:dataGen}

In the simulation study, observations $Y_i$ were drawn from the following model:

\begin{equation}
 Y_i \sim N(\mb X_i \bs \beta, \sigma^2).
 \label{eq:sim}
 \end{equation}

\(\mb X\) is the \(n \times p\) design matrix and \(\bs \beta\) is the
vector of coefficients of length \emph{p}; \(\sigma^2=1\) is the
residual variance. We investigate a one-dimensional scenario
(\emph{p}=1) and a high-dimensional scenario (\emph{p}=1,000). All
elements of \(\mb X\) were drawn independently from the standard normal
distribution. For the one-dimensional scenario, we consider the
following sample sizes \emph{n}: 20, 30, 50 and 100, and \(\beta\) is
set to either 0, 0.5, 1 or 1.5. For the high-dimensional scenario, we
consider sample sizes 30, 50, 75 and 100, and the first 10 entries of
\(\bs \beta\) are all set to either 0, 0.5 or 1; all other entries are 0. In
the one-dimensional scenario, 1,000 Monte-Carlo dataset instances are generated,
and in the high-dimensional scenario 100.

\subsubsection{\texorpdfstring{Prediction models
\label{sec:predModels}}{Prediction models }}\label{prediction-models}

In the one-dimensional scenario, the outcome was predicted using
ordinary least squares (OLS) using the \(lm.fit\) function in the
\emph{stats} R-package. In the high-dimensional scenario, the outcome was
predicted using elastic net (EN) \parencite{Zou2005} with fixed mixing
parameter 0.5 using the \emph{glmnet} function from the \emph{glmnet}
R-package \parencite{Friedman2010}. The penalty parameter was
tuned through an inner loop of 10-fold CV as implemented in the \emph{cv.glmnet} function.

\subsubsection{The MSE and its standard error} \label{sec:MSEest}

The out-of-sample MSE was estimated via either 10-fold CV as in \textcite{Bates2021}, or via the .632 bootstrap as in
\textcite{Efron1997}. Corresponding standard errors were calculated via nested CV with 9 inner folds \parencite{Bates2021} or via empirical influence functions \parencite{Efron1997}, respectively. The number of splits into cross-validation folds was repeated as recommended by \textcite{Bates2021}, with 1, 25, 100 or 200 splits in the one-dimensional scenario, or 5, 25 or 100 splits in the high-dimensional scenario. The number of bootstrap samples for .632 bootstrap estimation of the MSE was varied between 25, 100 and 200. For estimating the correlation between MSE and MST estimators through bootstrapping, either 10, 50, 100 or 500 bootstrap instances were used.

\subsubsection{True values and diagnostics}\label{diagnostics}

The true out-of-sample MSE is not known from the generative model \eqref{eq:sim} alone, as it depends on the accuracy of the estimation of $m$. Instead, it is approximated through Monte-Carlo simulation, by generating 5,000 datasets according to \eqref{eq:sim} with the same array of sample sizes $n$ and coefficients $\beta$ as in Section
\ref{sec:dataGen}, fitting the prediction model to these datasets and
evaluating their predictive performance on an independent test dataset drawn from
\eqref{eq:sim} with sample size 10,000. This yields 5000 precise estimates of the MSE; their average provides an approximation of the true out-of-sample MSE. The approximated true out-of-sample MSE in combination with the true MST given by $\frac{\sigma^2(n+1)}{n}$ is then used to approximate the true out-of-sample \rsq of this parameter setting and prediction method using \eqref{eq:R2oos} (see Tables
\ref{tab:trueR2} and \ref{tab:trueR2HD}). One-sided approximate z-tests were used to test whether \rsq $\leq 0$ in all scenarios with approximated true \rsq values below 0, and the proportion of times the null hypothesis was rejected was taken as an estimate of the type I error. A significance level of 5\% was used. Coverage of the 95\% confidence intervals is approximated as
the proportion of Monte-Carlo instances for which the confidence
interval includes the approximated true \rsq. In addition, the average
width of the confidence interval is calculated. 

The true SE of the \hrsq is approximated differently, as it also depends
on the data splitting algorithm and its settings (number of CV splits or
number of bootstraps). It is approximated by the standard deviation of
the \hrsq values of the algorithm and parameter settings concerned over all Monte-Carlo instances from Section
\ref{sec:dataGen}, see Supplementary Figures
\ref{fig:trueSER2}-\ref{fig:seMSE},
\ref{fig:trueSER2HD}-\ref{fig:trueSER2HDboot}. The true correlation \(\rho\) between \(\widehat{MSE}\) and \(\widehat{MST}\) was
approximated analogously as the empirical correlation of these
quantities over the same Monte-Carlo instances. The accuracy of the different
SE estimators was assessed by calculating the ratio of the estimated to
the approximated true SE, and taking the geometric mean over all
Monte-Carlo instances. Also the MSE of the estimated SE with respect to
the approximated true SE was calculated, which reflects both the
bias and variance of the estimators. The accuracy of the estimation of
\(\rho\) was assessed graphically using boxplots of the estimates
\(\hat{\rho}\). The whole pipeline of data generation, \rsq estimation and evaluation is shown in \fref{fig:flow}.

\tikzset{
blockPS2/.style={
    rectangle,
    draw,
	fill = purple!20,
    text width=27em,
    text centered,
    rounded corners
},
blockPS/.style={
    rectangle,
    draw,
	fill = purple!20,
    text width=38em,
    text centered,
    rounded corners
}
}
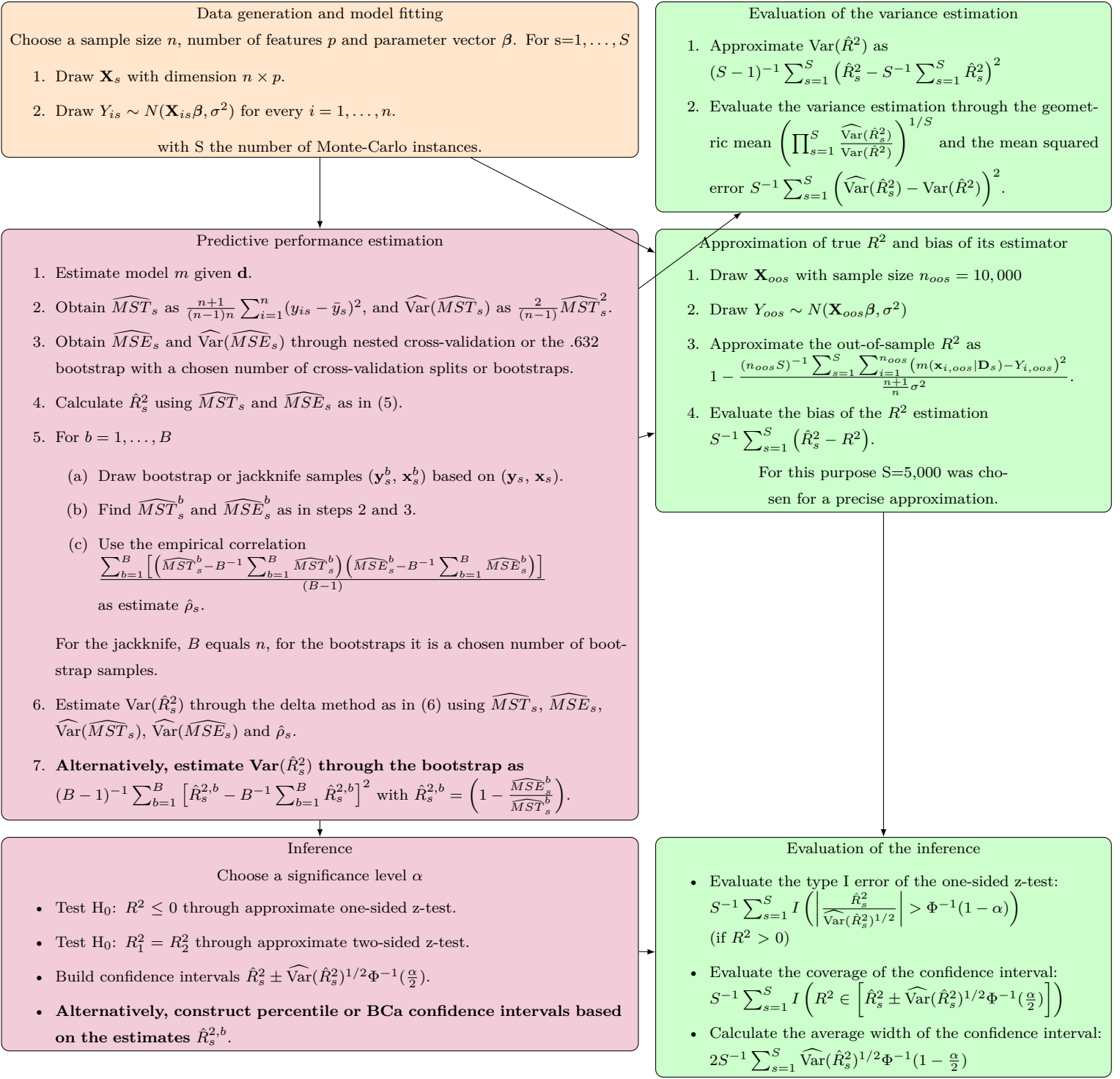
\begin{figure}
\vspace{-0.75cm}
\hspace{-1.9cm}
\begin{tikzpicture}\footnotesize
\matrix (m)[matrix of nodes, column sep=0.3cm, row sep=3mm, align=center, nodes={rectangle,draw, anchor=north} ]{
 |[blockPS][fill = orange!20,]| {Data generation and model fitting\\Choose a sample size $n$, number of features $p$ and parameter vector $\bs\beta$. For s=$1, \hdots, S$\\\begin{enumerate}
    \item{Draw  $\mb X_s$ with dimension $n\times p$.}
    \item{Draw $Y_{is} \sim N(\mb X_{is} \bs \beta, \sigma^2)$ for every $i=1,\hdots,n$.}
    \end{enumerate}with S the number of Monte-Carlo instances.}     &      
     |[blockPS2][fill = green!20,]| {Evaluation of the variance estimation\begin{enumerate}
 \item{Approximate $\Var{\hat{R}^2}$ as\\$(S-1)^{-1}\sum_{s=1}^S \left(\hat{R}^2_s - S^{-1}\sum_{s=1}^S \hat{R}^2_s \right)^2 $}
 \item{Evaluate the variance estimation through the geometric mean $\left(\prod_{s=1}^S \frac{\widehat{\text{Var}}(\hat{R}^2_s)}{\Var{\hat{R}^2}}\right)^{1/S}$ and the mean squared error $S^{-1}\sum_{s=1}^S \left(\widehat{\text{Var}}(\hat{R}^2_s) - \Var{\hat{R}^2}\right)^2$.}
 \end{enumerate}}\\ 
 |[blockPS]| {Predictive performance estimation\begin{enumerate}
         \item{Estimate model $m$ given $\mb d$.}
\item{Obtain $\widehat{MST}_s$ as $\frac{n+1}{(n-1)n}\sum_{i=1}^n(y_{is}-\bar{y}_s)^2$, and $\widehat{\text{Var}}(\widehat{MST}_s)$ as $\frac{2}{(n-1)}\widehat{MST}_s^2$.}
\item{Obtain $\widehat{MSE}_s$ and $\widehat{\text{Var}}(\widehat{MSE}_s)$ through nested cross-validation or the .632 bootstrap with a chosen number of cross-validation splits or bootstraps.}
\item{Calculate $\hat{R}^2_s$ using $\widehat{MST}_s$ and $\widehat{MSE}_s$ as in \eqref{eq:R2cv}.}
\item{For $b=1,\hdots,B$
\begin{enumerate}
\item{Draw bootstrap or jackknife samples ($\mb y_s^b$, $\mb x_s^b$) based on ($\mb y_s$, $\mb x_s$).}
\item{Find $\widehat{MST}_s^b$ and $\widehat{MSE}_s^b$ as in steps 2 and 3.}
\item{Use the empirical correlation\\$\frac{\sum_{b=1}^B \left[\left(\widehat{MST}_s^b-B^{-1}\sum_{b=1}^B\widehat{MST}_s^b\right)\left(\widehat{MSE}_s^b-B^{-1}\sum_{b=1}^B\widehat{MSE}_s^b\right)\right]}{(B-1)}$\\as estimate $\hat{\rho}_s$.}
\end{enumerate}
For the jackknife, $B$ equals $n$, for the bootstraps it is a chosen number of bootstrap samples.
}
\item{Estimate $\Var{\hat{R}^2_s}$ through the delta method as in \eqref{eq:delta} using $\widehat{MST}_s$, $\widehat{MSE}_s$, $\widehat{\text{Var}}(\widehat{MST}_s)$, $\widehat{\text{Var}}(\widehat{MSE_s})$ and $\hat{\rho}_s$.}
\item{\textbf{Alternatively, estimate $\Var{\hat{R}^2_s}$ through the bootstrap as}\\$(B-1)^{-1}\sum_{b=1}^B \left[\hat{R}^{2,b}_s-B^{-1}\sum_{b=1}^B \hat{R}^{2,b}_s\right]^2$ with $\hat{R}^{2,b}_s = \left(1-\frac{\widehat{MSE}_s^b}{\widehat{MST}_s^b}\right)$.} 
\end{enumerate}}  &
|[blockPS2][fill = green!20,]| {Approximation of true \rsq and bias of its estimator\begin{enumerate}
    \item{Draw  $\mb X_{oos}$ with sample size $n_{oos} = 10,000$}
    \item{Draw $Y_{oos} \sim N(\mb X_{oos} \bs \beta, \sigma^2)$}
    \item{Approximate the out-of-sample $R^2$ as\\$1-\frac{(n_{oos}S)^{-1}\sum_{s=1}^S\sum_{i=1}^{n_{oos}} \left(m(\mb x_{i, oos}\mid \mb{D}_s)-Y_{i,oos}\right)^2}{\frac{n+1}{n}\sigma^2}$.}
    \item{Evaluate the bias of the \rsq estimation $S^{-1}\sum_{s=1}^S \left(\hat{R}^2_s-R^2\right)$.}
    \end{enumerate}
    For this purpose S=5,000 was chosen for a precise approximation.
    }    \\
  |[blockPS]| {Inference\\Choose a significance level $\alpha$\begin{itemize}
\item{Test H$_0$: \rsq $\leq$ 0 through approximate one-sided z-test.}
\item{Test H$_0$: $R^2_1$ = $R^2_2$ through approximate two-sided z-test.}
\item{Build confidence intervals $\hat{R}_s^2 \pm \widehat{\text{Var}}({\hat{R}_s^2})^{1/2}\Phi^{-1}(\frac{\alpha}{2})$.}
\item{\textbf{Alternatively, construct percentile or BCa confidence intervals based on the estimates} $\hat{R}^{2,b}_s$.}
\end{itemize}}   & 
 |[blockPS2][fill = green!20,]| {Evaluation of the inference\begin{itemize}
 \item{Evaluate the type I error of the one-sided z-test: $S^{-1}\sum_{s=1}^S I\left(\left|\frac{\hat{R}_s^2}{\widehat{\text{Var}}({\hat{R}_s^2})^{1/2}}\right| > \Phi^{-1}(1-\alpha)\right)$\\(if \rsq > 0)}
 \item{Evaluate the coverage of the confidence interval: $S^{-1}\sum_{s=1}^S I\left(R^2 \in \left[\hat{R}_s^2 \pm \widehat{\text{Var}}({\hat{R}_s^2})^{1/2}\Phi^{-1}(\frac{\alpha}{2})\right]\right)$}
 \item{Calculate the average width of the confidence interval: $2S^{-1}\sum_{s=1}^S \widehat{\text{Var}}({\hat{R}_s^2})^{1/2}\Phi^{-1}(1-\frac{\alpha}{2})$}
 \end{itemize}}\\
 };
\path [>=latex,->] (m-2-1) edge (m-3-1);
\path [>=latex,->] (m-1-1) edge (m-2-2);
\path [>=latex,->] (m-1-1) edge (m-2-1);
\path [>=latex,->] (m-2-1) edge (m-2-2);
\path [>=latex,->] (m-3-1) edge (m-3-2);
\path [>=latex,->] (m-2-2) edge (m-3-2);
\path [>=latex,->] (m-2-1) edge (m-1-2);
\end{tikzpicture}
\caption{\label{fig:flow}Schematic overview of the simulation study. The top left panel (in orange) describes the repeated data generation. The center and bottom left panels (in purple) describe the estimation of \rsq and $\Var{\hat{R}^2}$ and statistical inference on either simulated or real data. The right panels (in green) describe the approximation of the true \rsq and $\Var{\hat{R}^2}$, as well as performance evaluation, which is only possible for simulated data where the ground truth is known. Arrows indicate input of one block into another. Bold text indicates competitor methods that depart from the methodology proposed in the paper.}
\end{figure}

\subsection{Results}

\subsubsection{The components of the \rsq: MST and MSE}

The unbiasedness of the MST estimator \eqref{eq:MST} is demonstrated numerically in \fref{fig:biasMST}. In the one-dimensional scenario, MSE estimation through .632 bootstrap is downward biased, whereas the cross-validation estimation with bias correction proposed by \textcite{Bates2021} is unbiased (Supplementary Figures \ref{fig:biasMSE} and \ref{fig:biasMSEHD}). The estimation of \Var{$\widehat{MSE}$} through CV or .632 bootstrap suffers from a small upward bias at small sample sizes, but this bias decreases as the sample size grows (Supplementary Figures \ref{fig:biasMSESE}-\ref{fig:biasMSESEboot}).

For the high-dimensional scenario, \sfref{fig:biasMSEHD} reveals an upward bias in MSE estimation for large sample sizes and large effect sizes. This bias results from the fact that in k-fold CV the model is trained on a dataset of size $n\frac{k-2}{k}$ in the nested CV scheme, rather than the models trained on sample size $n$ for which CV is attempting to estimate the performance. The bias correction proposed  in Appendix C of \textcite{Bates2021} worked fine in the one-dimensional case, as evident from \fref{fig:biasMSE}, but fails for the high-dimensional case, presumably because we are in the proportional, sparse regime with both $n$ and $p$ going to infinity, as mentioned by \textcite{Bates2021}. This bias could be reduced by increasing the number of folds \parencite{Bates2021}.

\subsubsection{Pooling versus averaging estimation of the \rsq}

The performance of the different alternative estimators of the \rsq is shown in \sfref{fig:altR2} for the one-dimensional scenario. The pooling \rsq is an unbiased estimator of the true \rsq as defined in \eqref{eq:R2oos} with low variance. The averaging \rsq with training MST also has a low variance, but is downward biased for weak signal strengths. The averaging \rsq with test MST estimator is very variable and dramatically downward biased for smaller sample sizes, and even at a sample size of 100 some of the bias persists. For this reason, we choose to work with the pooling estimator \eqref{eq:R2cv}. The normality assumption for this pooling estimator \hrsq, required for the delta method approximation of the SE and for construction of the confidence intervals, is assessed in Supplementary Figures \ref{fig:trueR2plot}-\ref{fig:trueR2plotHDboot}. Some departures from normality can be seen, especially at small sample sizes.

\subsubsection{One-dimensional scenario}

\begin{figure}
\centering
\includegraphics{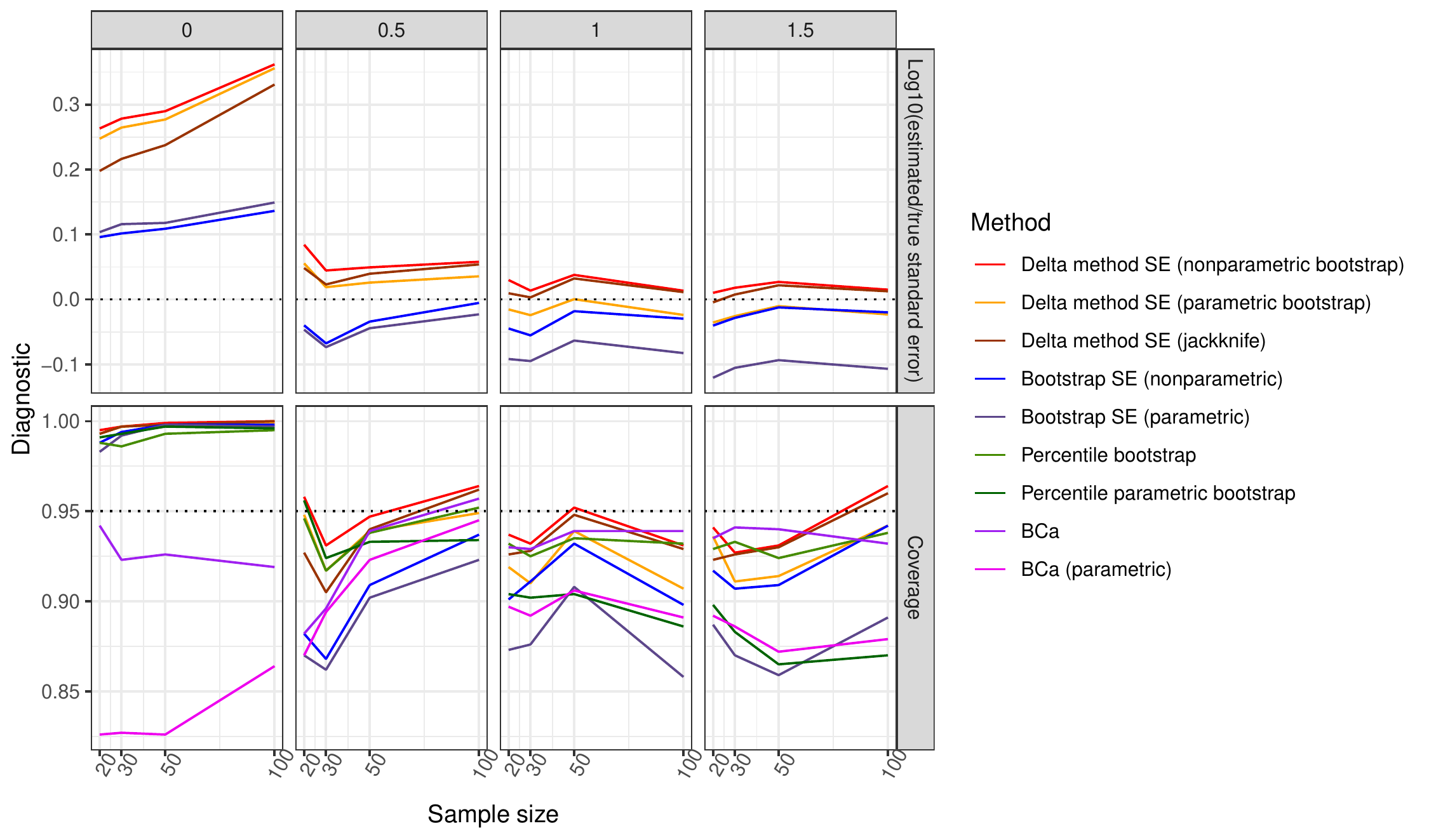}
\caption{Diagnostics for the one-dimensional simulation scenario using cross-validation: log10 of the geometric mean of the ratio of estimated to
approximated true standard error (SE) of the \rsq (top panels) and coverage of the confidence intervals (bottom panels) as a function of estimation method (colour), sample size (x-axis) and effect size (columns) for 500 bootstraps and 200 repeats of the cross-validation splits. The dotted lines indicate unbiased \rsq estimation and the nominal coverage of 95\%, respectively.\label{fig:OneDim}}
\end{figure}

The accuracy of the SE(\hrsq) estimation is shown in \fref{fig:OneDim} and \sfref{fig:mseUni} for cross-validation in the one-dimensional
scenario. The bias for delta method standard errors is mostly positive (i.e. conservative) and decreases with effect size. The nonparametric bootstrap method performs best for the estimation of the correlation \(\rho\) between $\widehat{MSE}$ and $\widehat{MST}$ (\sfref{fig:corEst}). Bootstrap standard errors are more accurate than delta method standard errors in
the null scenario, but tend to be downward biased (i.e. liberal) as the signal strength increases, especially for the parametric bootstrap. The coverage of the confidence intervals (\fref{fig:OneDim}) is close to the nominal level
for the delta method SE with nonparametric bootstrap or jackknife estimation of \(\rho\), but the intervals based on the bootstrap
(bootstrap SE, percentile and BCa) show undercoverage in some scenarios. When the predictors are not predictive of the outcome at all, all confidence intervals except the BCa intervals have a coverage above the nominal level of 95\%. The delta method SE confidence intervals are generally wider than the bootstrap confidence intervals (\fref{fig:sweepWidth}). All methods control the type I error of the approximate one-sided z-test below the significance level of 5\% (\fref{fig:zTest}). Similar conclusions are reached for the .632 bootstrap (Figures \ref{fig:biasUniBoot}-\ref{fig:zTestBoot}), even though it is slightly upward biased for \rsq estimation (\fref{fig:biasR2Uni}).

\begin{figure}
\centering
\includegraphics{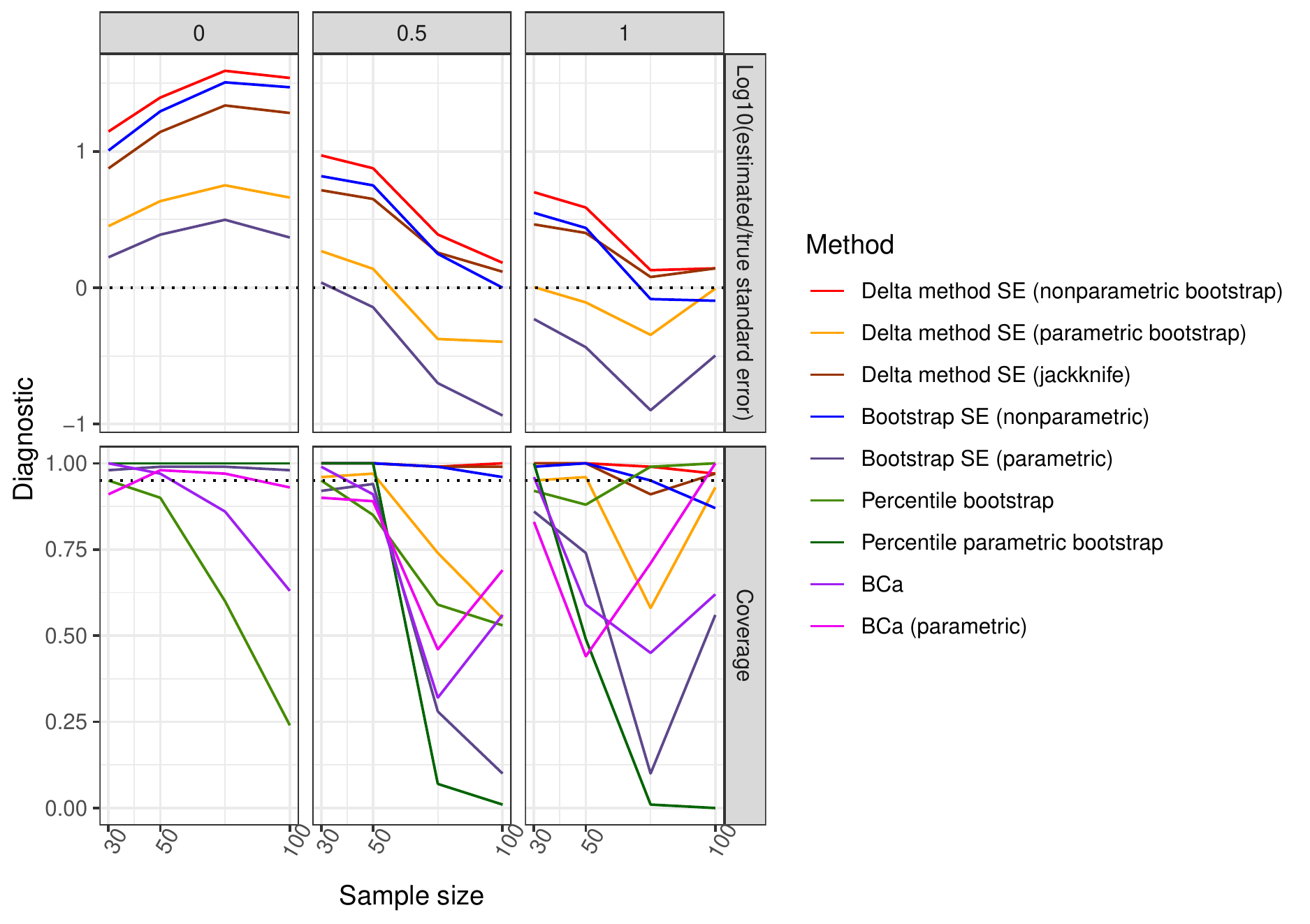}
\caption{Diagnostics for the high-dimensional simulation scenario using cross-validation: log10 of the geometric mean of the ratio of estimated to
approximated true standard error (SE) of the \rsq (top panels) and coverage of the confidence intervals (bottom panels) as a function of estimation method (colour), sample size (x-axis) and effect size (columns) for 100 bootstraps and 100 repeats of the cross-validation splits. The dotted lines indicate unbiased \rsq estimation and the nominal coverage of 95\%, respectively.\label{fig:HighDim}}
\end{figure}

\subsubsection{High-dimensional scenario}

In the high-dimensional scenario when using CV, the parametric bootstrap yields the best estimate of the SE of the \rsq in the null scenario of no predictive value, but underestimates the SE for stronger signal strengths, especially at large sample sizes
(\fref{fig:HighDim} and \sfref{fig:biasHigh}). As the sample size and effect size increase, the delta method SE
with nonparametric bootstrap or jackknife estimation of \(\rho\) and nonparametric bootstrap SE perform best, converging on the true value (\sfref{fig:corEstHD}). These findings are also reflected in the coverages of the confidence intervals, which lie above the nominal level of 95\% in the null scenario for all methods except the percentile bootstrap and BCa intervals. As the signal strength increases, only the delta method SE confidence intervals with \(\rho\) estimation using nonparametric bootstrap or jackknife and nonparametric bootstrap SE confidence intervals maintain a coverage close to the nominal level (\fref{fig:HighDim}), although the low coverage of the other methods is partly caused by the bias in the MSE estimation at high sample sizes mentioned above (Supplementary Figures \ref{fig:biasMSEHD}-\ref{fig:biasR2HD}). For every sample size, the confidence intervals are much wider than for the one-dimensional scenario (\sfref{fig:sweepWidthHD}). The type I error is not controlled at the significance level for the bootstrap SE methods (\sfref{fig:zTestHD}). Similar results are found for .632 bootstrap estimation of the \rsq (Supplementary Figures \ref{fig:biasHighBoot}-\ref{fig:zTestHDBoot}). 

In both one- and high-dimensional scenarios, the correlation between
\(\widehat{MSE}\) and \(\widehat{MST}\) is close to 1 when the predictors have no predictive value, but decreases and in some cases becomes negative as the effect size increases (Supplementary Figures \ref{fig:corEst}, \ref{fig:corEstBoot}, \ref{fig:corEstHD} and \ref{fig:corEstHDBoot}).
The negative correlations are likely an effect of the randomness in the design matrix, where designs with more variable predictors lead to more variance in the outcome (high \(\widehat{MST}\)), but also to better parameter estimates and hence better predictions (low \(\widehat{MSE}\)).

\section{Case study: predictability of Brassica napus and Zea mays phenotypes}


Gene expression and phenotypes were measured for 62 \emph{Brassica napus} plants
\parencite{DeMeyer2022}, and 60 \emph{Zea Mays} plants \parencite{Cruz2020}. For each crop, 5 phenotypes were considered for prediction: leaf 8 width, number of branches, number of leaves, root width and number of seeds for \emph{B.
napus} and leaf blade length, leaf blade width, husk leaf length, ear length and plant height for \emph{Z. mays}. The gene expression measurements were \emph{rlog} transformed prior to analysis
\parencite{Love2014}. Only the 5,000 genes with the highest expression were retained for model fitting. The outcome phenotypes were predicted from these 5,000 genes through EN with the same settings as in the simulation study (see Section \ref{sec:predModels}), and 10-fold CV with 100 repeats of the split into folds was used to estimate the corresponding MSE. The SE on the resulting \hrsq was calculated using the delta method as in Section \ref{sec:sersq} with jackknife estimation of \(\rho\). In addition, the SE was estimated using the nonparametric bootstrap with 50 bootstrap samples. A one-sided approximate z-test was
performed to test whether \rsq\(\leq0\), and confidence intervals were constructed. 

Estimated \rsq values, standard errors, p-values and confidence intervals of the \emph{B. napus} and \emph{Z. mays} data are shown in Tables \ref{tab:R2se} and \ref{tab:R2seZM}, respectively. For \emph{B. napus}, leaf 8 width and
number of seeds have an \rsq significantly different from 0 according to
the one-sided approximate z-test based on the delta method SE; for
\emph{Z. mays} blade 16 width is significant according to
this test. The bootstrap SE's are mostly smaller, as in the simulations, and yield narrower confidence
intervals and smaller P-values, but with similar conclusions of the corresponding significance tests.

\begin{table}[ht]
\centering
\begin{tabular}{|l|c|cccc|cccc|}
  \hline
 & \multirow{2}{*}{$\hat{R}^2$} & \multicolumn{4}{|c|}{Delta method} & \multicolumn{4}{|c|}{Bootstrap}\\
& & SE & P-value & 2,5\% & 97,5\% & SE & P-value & 2,5\% & 97,5\% \\
  \hline
  Leaf 8 width & 0.72 & 0.07 & 8.06e-25 & 0.58 & 0.86 & 0.09 & 4.06e-15 & 0.54 & 0.91 \\ 
  Total branch count & 0.11 & 0.17 & 2.61e-01 & -0.22 & 0.44 & 0.16 & 2.58e-01 & -0.22 & 0.43 \\ 
  Number of leaves & 0.19 & 0.25 & 2.29e-01 & -0.30 & 0.67 & 0.15 & 1.16e-01 & -0.12 & 0.49 \\ 
  Root system width & -0.02 & 0.17 & 5.46e-01 & -0.35 & 0.31 & 0.22 & 5.35e-01 & -0.45 & 0.41 \\ 
  Number of seeds & 0.42 & 0.20 & 1.77e-02 & 0.03 & 0.82 & 0.11 & 3.97e-05 & 0.21 & 0.63 \\ 
   \hline
\end{tabular}
\caption{\label{tab:R2se}Estimated \rsq and corresponding delta method and nonparametric bootstrap standard error (SE), one-sided p-value and lower and upper bound of the 95\% confidence interval for 5 \textit{Brassica napus} phenotypes.} 
\end{table}
\begin{table}[ht]
\centering
\begin{tabular}{|l|c|cccc|cccc|}
  \hline
 & \multirow{2}{*}{$\hat{R}^2$} & \multicolumn{4}{|c|}{Delta method} & \multicolumn{4}{|c|}{Bootstrap}\\
& & SE & P-value & 2,5\% & 97,5\% & SE & P-value & 2,5\% & 97,5\% \\
  \hline
  Blade 16 length & 0.30 & 0.23 & 9.60e-02 & -0.15 & 0.74 & 0.16 & 3.15e-02 & -0.02 & 0.61 \\ 
  Blade 16 width & 0.49 & 0.21 & 1.08e-02 & 0.07 & 0.90 & 0.09 & 1.44e-07 & 0.30 & 0.67 \\ 
  Husk leaf length & 0.30 & 0.29 & 1.49e-01 & -0.27 & 0.87 & 0.19 & 5.86e-02 & -0.08 & 0.68 \\ 
  Ear length & -0.02 & 0.18 & 5.41e-01 & -0.38 & 0.34 & 0.24 & 5.32e-01 & -0.48 & 0.44 \\ 
  Plant height & -0.01 & 0.15 & 5.36e-01 & -0.30 & 0.27 & 0.19 & 5.28e-01 & -0.38 & 0.35 \\ 
   \hline
\end{tabular}
\caption{\label{tab:R2seZM}Estimated \rsq, corresponding delta method and nonparametric bootstrap standard error (SE), one-sided p-value and lower and upper bound of the 95\% confidence interval for 5 \textit{Zea mays} phenotypes. For blade 16 length with the bootstrap SE, the p-value is a significant whereas the confidence includes 0; this is because the p-value is one-sided but the confidence interval is two sided.} 
\end{table}

\clearpage

A relevant scientific question is the comparison of the \rsq estimates for different phenotypes
within the same dataset. If the design matrix were fixed, the different
estimates would be independent and the variance of the difference would simply
equal the sum of the variances of both \rsq estimates. However, if the design
matrix is random, as is the case here with the gene expression
measurements, the MSE estimates of two phenotypes correlate. The variance on the estimator for the difference between the \rsq of two different phenotypes $a$ and $b$ then equals $\Var{R^2_a - R^2_b} = \Var{R^2_a} + \Var{R^2_b} - 2 \Cor{R^2_a ,R^2_b}\sqrt{\Var{R^2_a}\Var{R^2_b}}$. The correlation $\Cor{R^2_a ,R^2_b}$ was estimated by the bootstrap: samples from the gene expression matrix were sampled with replacement 50 times, and corresponding entries of the phenotypes $a$ and $b$ are used to estimate the \rsq. The empirical correlation of these 50 bootstrap estimates is then used as an estimate of $\Cor{R^2_a ,R^2_b}$. The test statistics and p-values using the delta method SE for the approximate two-sided z-test
of the null hypothesis of zero difference between the \rsq values for \emph{B. napus} are
shown in Table \ref{tab:R2diffZ}. The leaf 8 width is the only phenotype with an \hrsq significantly different from some other phenotypes: total branch count, number of leaves and root system width. 

\begin{table}[ht]
\centering
\begin{tabular}{l|cccc}
  \hline
 & Leaf 8 width & Total branch count & Number of leaves & Root system width \\ 
  \hline
  Total branch count & -2.98 (0.0029) &  &  &  \\ 
  Number of leaves & -2.39 (0.017) & 0.27 (0.79) &  &  \\ 
  Root system width & -2.45 (0.014) & -0.39 (0.7) & -0.61 (0.54) &  \\ 
  Number of seeds & -1.59 (0.11) & 1.69 (0.091) & 0.88 (0.38) & 1.41 (0.16) \\ 
   \hline
\end{tabular}
\caption{\label{tab:R2diffZ}Approximate two-sided z-statistic for difference in \rsq of \textit{Brassica napus} phenotypes in columns and rows, with corresponding p-value calculated using the delta method SE between brackets. For instance, the top left entry is the total branch count \rsq minus leaf 8 width \rsq, divided by a standard error estimate for this difference.}
\end{table}

A further research question could be whether the predictability of certain
phenotypes differs between species. Since they are calculated on
independent datasets, the variance of the difference between two
\rsq estimators is simply the sum of the variances. Hence the approximate two-sided z-statistic
for the difference between the \rsq values of, e.g.~\emph{B. napus} leaf
8 width and \emph{Z. mays} blade leaf width is, using the delta method
SE's, \(\frac{0.72-0.49}{\sqrt{0.07^2+0.21^2}} = 1.06\), so not
significant (p-value = \(0.28\)). Yet the standardized difference in
predictability between \emph{B. napus} leaf 8 width and \emph{Z. mays}
plant height \(\frac{0.72-(-0.01)}{\sqrt{0.07^2+0.15^2}} = 4.52\) is
indeed significant (p-value = \(\ensuremath{6.2\times10^{-6}}\)).

\hypertarget{discussion}{%
\section{Discussion}\label{discussion}}

Research into \rsq-like measures has generally focused on in-sample
performance, yet the \rsq is also frequently employed to score
out-of-sample prediction. Here we have formally defined the
out-of-sample \rsq as one minus the ratio of prediction loss of a particular
prediction model to the prediction loss of the null model ignoring
covariate information. The resulting \rsq then has a clear interpretation: when it is larger than 0, the
prediction model is useful for out-of-sample prediction. When it is
smaller than 0, the average of the observed data is a better predictive
model, either because the predictors do not contain enough information
on the outcome vector, because the sample size is too small to allow for
accurate model fitting, or because the model is ill-suited to the prediction task. We demonstrated how the out-of-sample \rsq can be estimated using data splitting algorithms. We found that the pooling \rsq estimator, which separately estimates the squared error losses of the null and prediction models and only then combines them into a final estimate \hrsq, is unbiased. Hence this pooling estimator should be preferred to averaging estimators that calculate \hrsq values in every cross-validation fold separately and then average over the folds, which suffer from bias. Unlike the .632 bootstrap, cross-validation in combination with the pooling estimator provides almost unbiased estimates of the \rsq, in agreement with previous findings \parencite{BragaNeto2004, Jiang2007, Kohavi2001, Molinaro2005}, and should be the preferred way to estimate predictive performance. Yet some downward estimation bias appears in high-dimensional settings with strong signal, but this can be countered by a sufficiently high number of folds if computationally feasible. Finally, it is good to remember that the \rsq estimate does not apply specifically to the model trained on the dataset at hand, but rather to all models trained on similar, randomly drawn datasets. 

Like any parameter estimated from a finite sample, estimates obtained
through data splitting algorithms are uncertain. Yet by lack of
estimators for \rsq-like measures' standard errors, often only their
point estimates are reported. These alone do not allow simple
statistical questions to be answered, such as whether the predictive
model significantly reduces the loss with respect to the null model. One way to answer this question is by
repeatedly permuting the outcome variable, and each time refitting the
predictive model and calculating predictive performance, thus building a
null distribution of the performance estimate \parencite{Cruz2020, DeMeyer2022}. Such permutation methods
have the downside of being computationally demanding, and of not stating
a clear null value for the \rsq. The null hypothesis tested by these permutation methods is that
the predictors are not predictive of the outcome, which implies some
unknown \rsq value below 0 (see Supplementary Section \ref{sec:R2null}). Also,
permutation methods only yield p-values, but no standard errors.

As an alternative, we have provided a standard error for the
out-of-sample \rsq estimated through cross-validation or the .632 bootstrap
by building on recent advances in standard error estimation for the
mean squared error loss. Our method is also easily extendable to
\rsq values estimated on independent test data. The standard errors on \hrsq
allow for testing the null hypothesis H$_0$: \(R^2 \leq 0\), which mirrors
the interpretable alternative hypothesis H$_a$: \(R^2 > 0\) that indicates
that the predictive model significantly improves upon the null model without predictors. The standard error on \hrsq can also be used for the construction of confidence intervals and for comparing predictability
of different outcome variables with possibly different units. Comparison of different prediction models for the same outcome variable, on the other hand, is better done directly on the MSE values, as this does not require the approximation of the variance of a ratio through the delta method.

The delta method standard error estimators we provided are upward biased in some
settings. Possible explanations are a poor approximation by the delta method at low sample sizes when the departure from normality of the estimator \hrsq is strongest (see Supplementary section \ref{sec:R2dist}), and the difficulty in estimating the correlation $\rho$ between $\widehat{MSE}$ and $\widehat{MST}$. Yet the delta method standard errors, using nonparametric or jackknife estimates of $\rho$,  are still preferable to bootstrap standard errors, which can be
downward biased and lead to loss of type I error control and lower than nominal coverage of the confidence intervals. The variance of the estimator of the out-of-sample \rsq is small when there is hardly any predictive value in the predictors, but all methods considered overestimate this variance. Fortunately, this bias decreases as the predictive value increases, promising good power to detect truly predictable outcomes. Nevertheless, the estimator variance of \hrsq was found to be considerable in the high-dimensional scenario, supposedly because of the variability in the model fitting of high-dimensional, penalized models. This cautions against overinterpretation of (subtle differences between) \hrsq values estimated for such models. Hence, we encourage reporting standard errors and confidence intervals for diagnostics of predictive models to provide insight into the reproducibility of the result, guide follow-up study design, and allow for model comparison.

\bigskip
\begin{center}
{\large\bf SUPPLEMENTARY MATERIAL}
\end{center}

\begin{description}

\item[Supplementary material] Exhaustive simulation results, proofs and software versions. (pdf)
\item[R-code] R-code for calculation of the \rsq and its standard error, and for running all simulations and analyses is available at \url{https://github.com/maerelab/Rsquared.} (url)

\end{description}

\printbibliography

\end{document}